\documentclass[12pt,preprint]{aastex}

\def\etal{{\it et al.}}
\def\eg{{\it e.g.,~}}
\def\ie{{\it i.e.,~}}
\def\tcl{t_{\rm cool}}
\def\lcl{l_{\rm cool}}
\def\kms{~{\rm km~s^{-1}}}
\def\cm3{~{\rm cm^{-3}}}
\def\kpc{~{\rm kpc}}
\def\Msun{~{\rm M}_{\sun}}

\slugcomment{draft of \today}
\shorttitle{Thermal-Gravitational Instability}
\shortauthors{Baek et al.}

\begin{document}

\title{Effects of Rotation on Thermal-Gravitational Instability
in the Protogalactic Disk Environment} 

\author{Chang Hyun Baek\altaffilmark{1,2}, Dongsu Ryu\altaffilmark{3}, 
Hyesung Kang\altaffilmark{4}, and Jongsoo Kim\altaffilmark{2}}

\altaffiltext{1}{ARCSEC, Sejong University, Seoul 143-747, Korea}
\altaffiltext{2}{Korea Astronomy \& Space Science Institute, Daejeon
                 305-348, Korea}
\altaffiltext{3}{Department of Astronomy \& Space Science, Chungnam National
                 University, Daejeon 305-764, Korea}
\altaffiltext{4}{Department of Earth Sciences, Pusan National University,
                 Pusan 609-735, Korea}

\begin{abstract}

Thermal-gravitational instability (TGI) is studied in the protogalactic
environment.
We extend our previous work, where we found that dense clumps first
form out of hot background gas by thermal instability and later a
small fraction of them grow to virialized clouds of mass 
$M_c \ga 6\times10^6 \Msun$ by gravitational infall and merging.
But these clouds have large angular momentum, so they would be
difficult, if not impossible, to further evolve into globular clusters.
In this paper, through three-dimensional hydrodynamic simulations 
in a uniformly rotating frame,
we explore if the Coriolis force due to rotation in protogalactic disk
regions can hinder binary merging and reduce angular momentum of the
clouds formed.
With rotation comparable to the Galactic rotation at the Solar circle,
the Coriolis force is smaller than the pressure force during the early
thermal instability stage.
So the properties of clumps formed by thermal instability are not
affected noticeably by rotation, except increased angular momentum.
However, during later stage the Coriolis force becomes dominant over
the gravity, and hence the further growth to gravitationally bound
clouds by gravitational infall and merging is prohibited.
Our results show that the Coriolis force effectively destroys the 
picture of cloud formation via TGI, rather than alleviate the problem 
of large angular momentum.

\end{abstract}


\section{Introduction}
Thermal instability (TI) \citep{fie65} is often invoked to explain
a variety of physical phenomena in astrophysical environments:
for instance, the multiple phases of interstellar gas
\citep[\eg][]{fgh69,mo77}, the formation of globular clusters
\citep[\eg][]{fall85,kang00}, cooling flows in clusters of galaxies
\citep[\eg][]{nuls86}, and the generation of turbulent flows in
the interstellar medium \citep[\eg][]{vaz00,kri02}.
Although the basic concept of TI as a local instability is rather
simple and robust, the realistic situation is often more complex,
involving effects such as magnetic field, turbulence, gravity, and
rotation in case of galactic disks.

Recently the formation of structures via thermal-gravitational
instability (TGI) in the protogalactic halo environment was studied
using three-dimensional numerical simulations
\citep[hereafter Paper 1]{bkkr05}. 
The growth of density perturbations initially via TI and subsequently
via gravitational infall and merging was followed up to $16-20$ cooling
times in a periodic cubic box with size $L=$ 10 kpc. 
The simulations showed that clumps emerge first on scales smaller
than the cooling length as a result of non-linear behavior of TI.
Those clumps grow through compression by background pressure,
as well as through gravitational infall. 
Later during the gravitational merging stage, some clumps become
gravitationally bound, virialized clouds with mass
$M_c\ga 6\times 10^6 M_{\sun}$ and radius $R_c\approx 150-200 $pc.
However, these massive clouds acquire angular momentum through tidal
torque and merging and have a large spin parameter
$\left<\lambda_s\right> \sim 0.3$.
Hence removal of the angular momentum from theses clouds is critical, 
if they were to collapse further to form halo globular clusters such as
in the model by \citet{fall85}.

In this paper we study the effects of rotation in protogalactic disk
regions on the formation of clouds via TGI.
Uniform rotation was included in the same numerical simulations as in
Paper 1 to model protogalactic disk environment.
In the next section we describe our models and numerical method.
The simulation results are presented in \S III.
Summary follows in \S IV.

\section{Numerical Simulations}
As in Paper 1, we consider a primordial gas of $T_h = 1.7\times 10^6$K,
which corresponds to the canonical temperature of an isothermal sphere 
with circular velocity $V_c=220\kms$, representing the hot phase of
the gas in disk galaxies like the Milky Way.
The fiducial value of the mean background density of hydrogen nuclei
was chosen to be $n_h = 0.1 \cm3$.
For the primordial gas with an assumed ratio of He/H number densities
of 1/10, the gas mass density is given by
$\rho_h = (2.34\times 10^{-24}~{\rm g})~n_h$.
With $T_h = 1.7\times 10^6$K and $n_h = 0.1 \cm3$, the initial cooling
time scale is $\tcl=2\times 10^7$ yrs.
On the other hand, the free-fall time scale, or the gravitational time
scale, is $t_{\rm grav} = 1.4 \times 10^8$ yrs $\approx 7\ \tcl$.
Note that $\tcl \propto n_h^{-1}$, while $t_{\rm grav} \propto n_h^{-1/2}$.
So cooling, compared to gravitational processes, becomes relatively more
important at higher densities and {\it vice versa}.
The cooling length scale is defined as
$\lcl = c_h \cdot t_{\rm cool} = 4$ kpc, where $c_h=198 \kms$ is
the sound speed.

To model the protogalactic disk environment, uniform rotation was
incorporated by using a Cartesian coordinate system that rotates with
angular velocity, $\Omega \hat z$.
We chose as a fiducial value for the angular speed
$\Omega_o =27\kms\kpc^{-1}$, which corresponds to the Galactic rotation
at the Solar circle ($R_o \approx 8.5 \kpc$) \citep{fw97}.
A case with lower the angular speed, $\Omega_o / 2$, was also considered
for comparison.
In the rotating frame, an additional Coriolis force term,
$f_c = -2 \Omega \hat {z} \times \mathbf v$, is added to the equation of
motion.
The simulation box was set to be cubic, with size $L = 10 \kpc = 2.5 \lcl$.
The size was chosen to be large enough to accommodate a fair number of
thermally unstable clouds of cooling length size, and so to obtain good
statistics of cloud properties.
Period boundary condition was imposed on the box, although it might
not be the most natural choice for a disk-like geometry.
We believe this particular choice of boundary condition would not
affect the main conclusion regarding the role of rotation. 
Our periodic, cubic, simulation box represents a volume of disk
region with significant rotation inside a protogalaxy. 

To mimic density perturbations existed on a wide range of length scales
inside the protogalaxies, the initial density field was drawn from
random Gaussian fluctuations with predefined density power spectrum.
The density power spectrum was assumed to be given by $P_k \propto k^n$
with $n=0$ (white noise).
In Paper 1 we showed that the properties of clouds, once formed, do not
depend on initial fluctuation spectrum, although their spatial distribution
is sensitive to the spectrum.
Without {\it a priori} knowledge on initial fluctuations, a spectrum of 
constant power over all scales was chosen.
The amplitude of the density power spectrum was fixed by the condition
$\delta_{\rm rms} \equiv \left<\delta \rho^2 \right>^{1/2}/\left<\rho\right>
= 0.2$. The initial temperature was set to be uniform, and the initial
velocity was set to be zero everywhere.

The evolution of the gas from the initial perturbations was followed with
1) radiative cooling due to the primordial gas \citep{suth93} down to
$T = 10^4$ K, 2) background heating equal to the cooling of the initial,
unperturbed background gas, and 3) self-gravity.
The hydrodynamics was solved using an Eulerian hydrodynamic code
based on the total variation diminishing scheme \citep{ryu93} on a grid
with $1024^3$ cells (or $512^3$ cells in the comparison run with
$\Omega_o / 2$).
Simulations started at $t = 0$ and lasted up to $t_{\rm end} =16 - 20 \tcl$.
Three simulations are presented in this paper, differing in angular speed.
Model parameters are summarized in Table 1.

\section{Results}

We begin our discussion by comparing the Coriolis force to pressure
force and gravity for the flows associated with the formation of clouds.
First, the ratio of the Coriolis force to pressure force can be 
estimated roughly as
\begin{equation}
 {f_c \over f_p} 
 \sim { {2 \Omega_o v} \over {P_h/R_c\rho_h}} 
 \sim { {10 \Omega_o R_c} \over 3 {c_h}},
\end{equation}
where the typical flow speed is assumed to be similar to the sound speed, 
\ie $v\sim c_h$, and the adiabatic index of gas is $\gamma=5/3$.
Here $R_c$ is the radius of typical clouds.
Conservatively with $R_c \sim 100$ pc, the ratio is $f_c/f_p \sim 0.05$.
So we expect that the effects of rotation are small, if not negligible,
during the TI stage.
The ratio of the Coriolis force to gravity can be written roughly as
\begin{equation}
 {f_c \over f_g} \sim { {2\Omega_o v R_c^2} \over {G M_c}}.
\end{equation}
Again with $v\sim c_h$, and conservatively with $M_c \sim 10^6\Msun$ and
$R_c \sim 100$ pc, the ratio becomes $f_c/f_g \sim 25$.
Hence during the gravitational infall and merging stages, the
Coriolis force is expected to play a dominant dynamical role in our
simulations with rotation.

The Coriolis force causes flows to be deflected at the right angle to the
flow direction, hence it hinders spherical infall motions toward high
density peaks, and generates circular motions in the plane perpendicular
to the rotation axis.
The Coriolis force also hinders the mergers of two clumps by drifting them
to the opposite directions perpendicular to the line connecting two clouds,
which leads to stretched worm-like structures.
As a result, the formation of knot-like structures is suppressed and
instead filamentary and sheet-like structures seem to appear.

Figure 1 shows the density power spectrum at different times.
In the figure the dimensionless wavenumber is given as $k \equiv L/\lambda$.
The power spectrum is presented in a way that
$\int P_k dk = \left< \rho^2 \right>$.
During early stage $t\la 6\tcl$, the evolution of the density power
spectrum looks similar in two models S1024 and R1024.
It is because with $t_{\rm cool} < t_{\rm grav}$, initially the power
grows mostly due to TI, and the Coriolis force has negligible effects.
During $t \ga 6\tcl$, gravity controls the growth and so the two models
evolve differently. 
In model S1024 the power continues to grow due to gravitational infall
and merging over all scales. 
In model R1024, on the other hand, the Coriolis force suppresses the
growth, more evidently on smaller scales.
The comparison of the power spectrum of models R1024 ($\Omega=\Omega_o$)
and RH512 ($\Omega=0.5\Omega_o$) demonstrates the effects of different
rotation speed.
The figure shows that even with a half rotation speed the Coriolis force
suppresses the growth of the power spectrum effectively.

To look at the shape and spatial distribution of formed structures,
three-dimensional isodensity surfaces of a cubic region are shown in
Figure 2.
Initially clumps appear via TI in both models, as discussed in Paper 1
(see Figure 3 there).
In model S1024, by $t=8\tcl$ the clumps have developed into distinct
clouds with roughly
spherical shapes.
By $t = 12 \tcl$ some clouds have grown to be massive enough to become
gravitationally bound, and by $t = 16 \tcl$ they have grown more massive.
Those gravitationally bound clouds have the central density higher than
1000 times the mean background density.
In model R1024, on the other hand, the initial clumps do not grow to
distinct clouds.
Instead loosely connected filamentary and sheet-like structures appear
with the central density lower than 100 times the mean background density. 

Although clouds are not distinctively defined in model R1024, we still
identified clouds around high density peaks by the algorithm
{\tt clumpfind} described in Paper 1, and calculated their properties.
The first row of Figure 3 shows the number of identified clouds, $N_c$,
as a function of their mass, $M_c$.
The mass function of the two models is almost identical at $t=4 \tcl$.
It is roughly Gaussian, since the initial density perturbations were
drawn from a random Gaussian distribution. 
In model S1024, the mass function has evolved roughly into a log-normal
distribution by $t = 8 \tcl$, which is a signature of nonlinear structure
formation.
At later stage the mass function extends to higher mass with
a power-law distribution, as more massive clouds develop through
gravitational merging.
In model R1024, however, the mass distribution remains roughly Gaussian at
$t = 8 \tcl$, and later it develops into a form, which is not well defined.
This is another indication that the development of nonlinear structures
has been severely altered by the Coriolis force.

The second row of Figure 3 shows the energy ratio of identified clouds, 
$\beta = {2 (E_K+E_T)}/|E_G|$.
Here $E_K$ is the kinetic energy defined in the center of mass of a given 
cloud, $E_T$ is the thermal energy, and $E_G$ is the gravitational energy.
The parameter $\beta$ tells us whether clouds are gravitationally bound
($\beta \la 2$), and whether in virial equilibrium ($\beta \sim 1$).
The figure shows that the virialized clouds with
$M_c \ga 6 \times 10^6\Msun$ have formed in model S1024.
But in model R1024 the identified objects have $M_c \la 10^6 \Msun$,
and none are gravitationally bound.
We note that even in model RH512 with a half rotation speed, no
gravitationally bound clouds were found.

The bottom row of Figure 3 shows the specific angular momentum of
identified clouds.
At $t=4 \tcl$ the specific angular momentum is different in the two
models, although other properties are similar.
As mentioned earlier, clumps formed in the early stage gain angular
momentum through the Coriolis effect, so $j_c$ is higher in model R1024
than in model S1024. 
But during later stage angular momentum is acquired efficiently
through torque and merging in model S1024, while it is not in model R1024.
Hence for a given mass, $j_c$ becomes higher in model S1024 than in
model R1024.

\section{Summary}

We study the effects of rotation in protogalactic disk regions on the
formation of structures via TGI in the protogalactic environment.
A simplified setting was considered, where a gas of primordial composition
evolves from initial density perturbations in a uniformly rotating box.
In Paper 1, we found that without rotation, virialized clouds of mass
$M_c \ga 6\times10^6 \Msun$ can form as a result of TGI although they form
with large angular momentum of spin parameter
$\left<\lambda_s\right> \sim 0.3$.
In this paper we find that with rotation whose angular speed is
comparable to that of the Galactic rotation at the Solar circle,
$\Omega_o =27\kms\kpc^{-1}$, the Coriolis force suppresses gravitational
infall and merging and disperses the gas to filamentary and sheet-like 
structures.
As a result, instead of massive virialized clouds formed in non-rotating
models, clumps with $M_c \la 10^6 \Msun$, which are gravitationally
unbound and often transient, are only found.

We conclude that the rotation in protogalactic disk regions has destructive
effects on the formation of clouds, rather than alleviate the angular
momentum problem discussed in Paper 1.
The results in this paper and those in Paper 1 combined imply that it
would be difficult for globular clusters to have formed via TGI in
protogalaxies, and even less likely in rotating disk regions.

A few notes on our results:
1) Primordial gas is considered in this paper.
But adding metal of order of 0.1 $Z_\sun$, which would enhance thermal
processes, does not change the results, because it is the gravitational
processes that are responsible for the formation of massive clouds but
suppressed by the Coriolis force.
We confirmed it with another simulation with metalicity of 0.1 $Z_\sun$,
although we do not report the simulation in the paper.
We point, however, that changing other parameters, for instance
increasing the gas density by a factor of 10 in particular, would
make differences.
2) The size of the simulation box we used, $L = 10 \kpc$, along with
periodic boundaries is too large to represent a protogalactic disk region.
In order to check the effects of box size, we performed simulations
in a smaller box of $2.5 \kpc$ with $256^3$ cells with and without
rotation.
Again although we do not report the simulation in details,
the results confirmed that the formation of gravitationally bound
objects is prohibited by the Coriolis force, regardless of the box size.
3) To be more realistic, differential rotation instead of uniform
rotation needs to be investigated.

\acknowledgments{
We thank the anonymous referee for clarifying comments.
The work of C.H.B., H.K. and J.K. was supported by KOSEF through 
Astrophysical Research Center for the Structure and Evolution of Cosmos 
(ARCSEC). The work of D.R. was supported by a Korea Research Foundation
grant (KRF-2004-015-C00213). Simulations were run using Linux clusters
at KASI and KISTI Supercomputing Center.

\clearpage

\begin{deluxetable}{cccc}
\tablewidth{0pc}
\tablecaption{Model Parameters for Simulations}
\tablehead{\colhead{Model} & \colhead{No. of Grid Zones}
& \colhead{$t_{\rm end}$ ($\tcl$)\tablenotemark{a}}
& \colhead{$\Omega$\tablenotemark{b}} }
\startdata
S1024 & $1024^3$ & 16 & 0.0 \\
R1024 & $1024^3$ & 16 & $\Omega_o$ \\
RH512 & $512^3$  & 20 & $0.5\Omega_o$ \\
\enddata
\tablenotetext{a}{$\tcl = 2\times 10^7$ yrs} 
\tablenotetext{b}{$\Omega_{0} = 27\kms \kpc^{-1}$.}
\end{deluxetable}

\clearpage

\begin{figure}
\includegraphics[width=18cm]{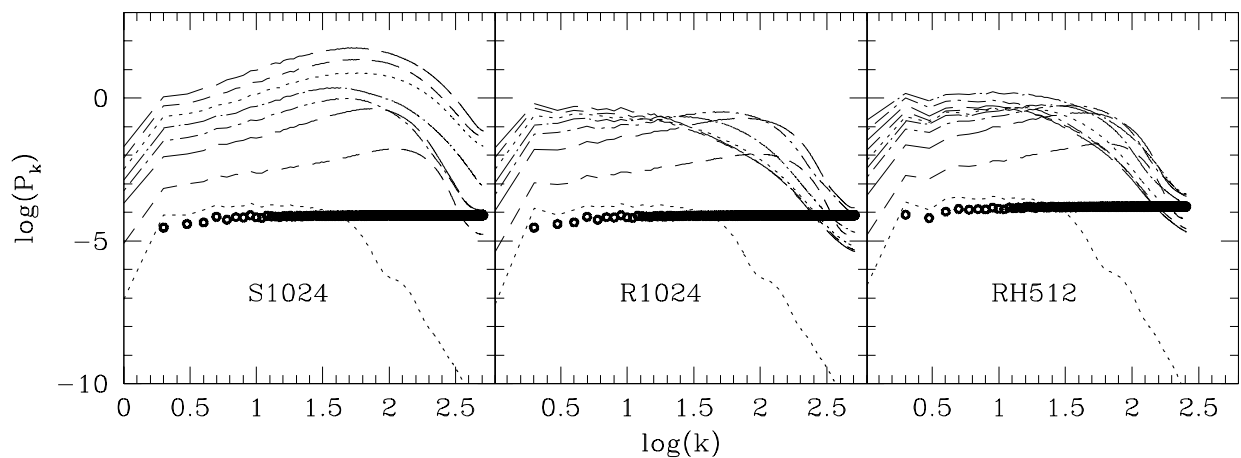}
\figcaption
{Evolution of the density power spectrum in models S1024, R1024, and RH512.
Circles represent the initial power spectrum at $t = 0$.
Lines show the power spectrum at $2~\tcl$, $4~\tcl$, $6~\tcl$, $\ldots$,
$16~\tcl$ in models S1024 and R1024, and at $4~\tcl$, $6~\tcl$, $\ldots$,
$20~\tcl$ in model RH512.}
\end{figure}

\begin{figure}
\plotone{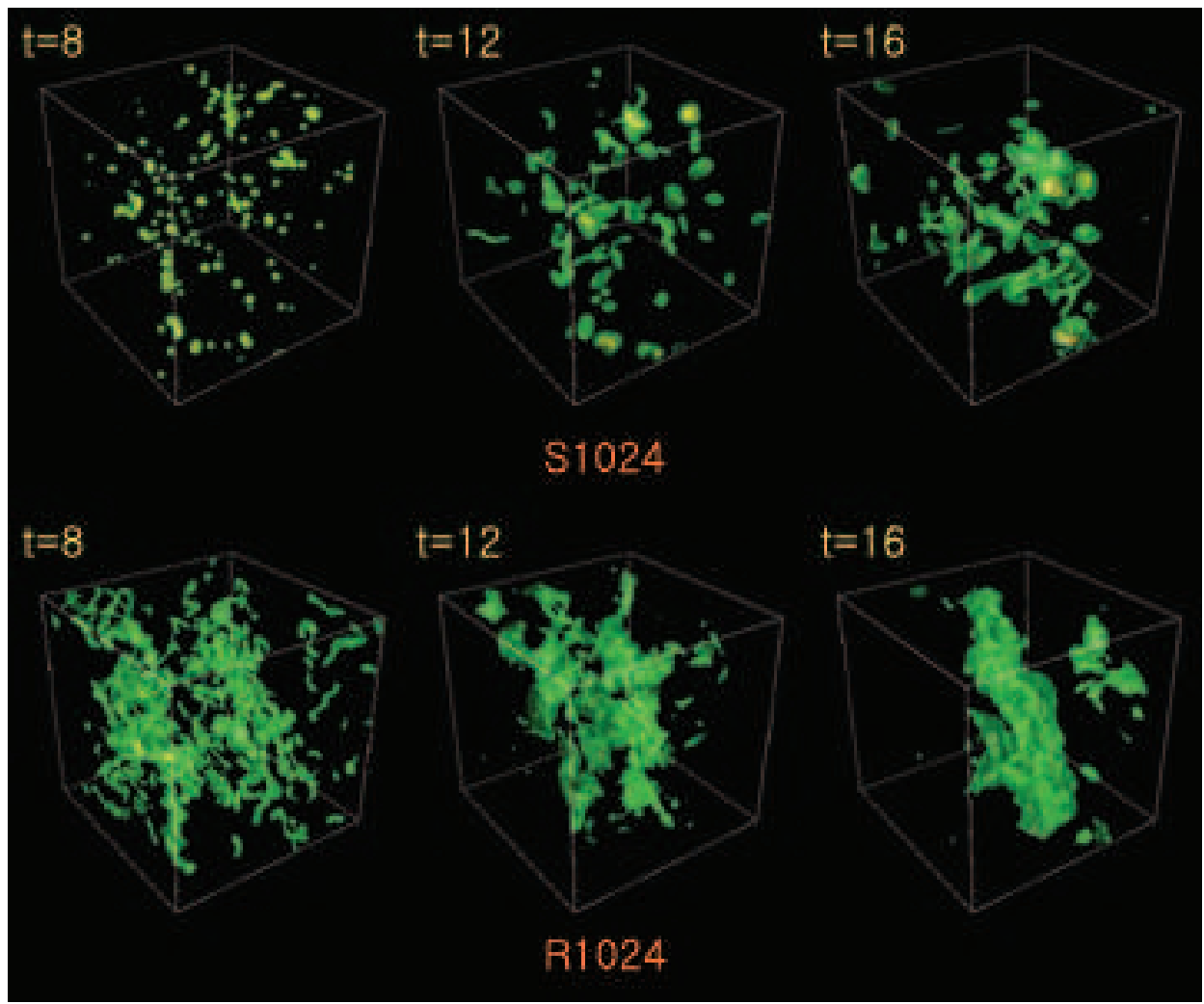}
\figcaption
{Isodensity surfaces inside a $2.5 \kpc$ box of $256^3$ grid zones,
which is $(1/4)^3$ of the total simulation box, at $8~\tcl$, $12~\tcl$,
and $16~\tcl$ in model S1024 (top panels), and in model R1024
(bottom panels).
Green surfaces correspond to $10\rho_0$, yellow surfaces to $10^2\rho_0$,
and red surfaces to $10^3 \rho_0$. 
Here $\rho_0$ is the mean initial density.
The box is oriented in such a way that the $x$, $y$, and $z-$axes are
from near to far, from bottom to top, and from left to right, respectively.
Rotation is along the $z-$axes in model R1024.}
\end{figure}

\begin{figure}
\plotone{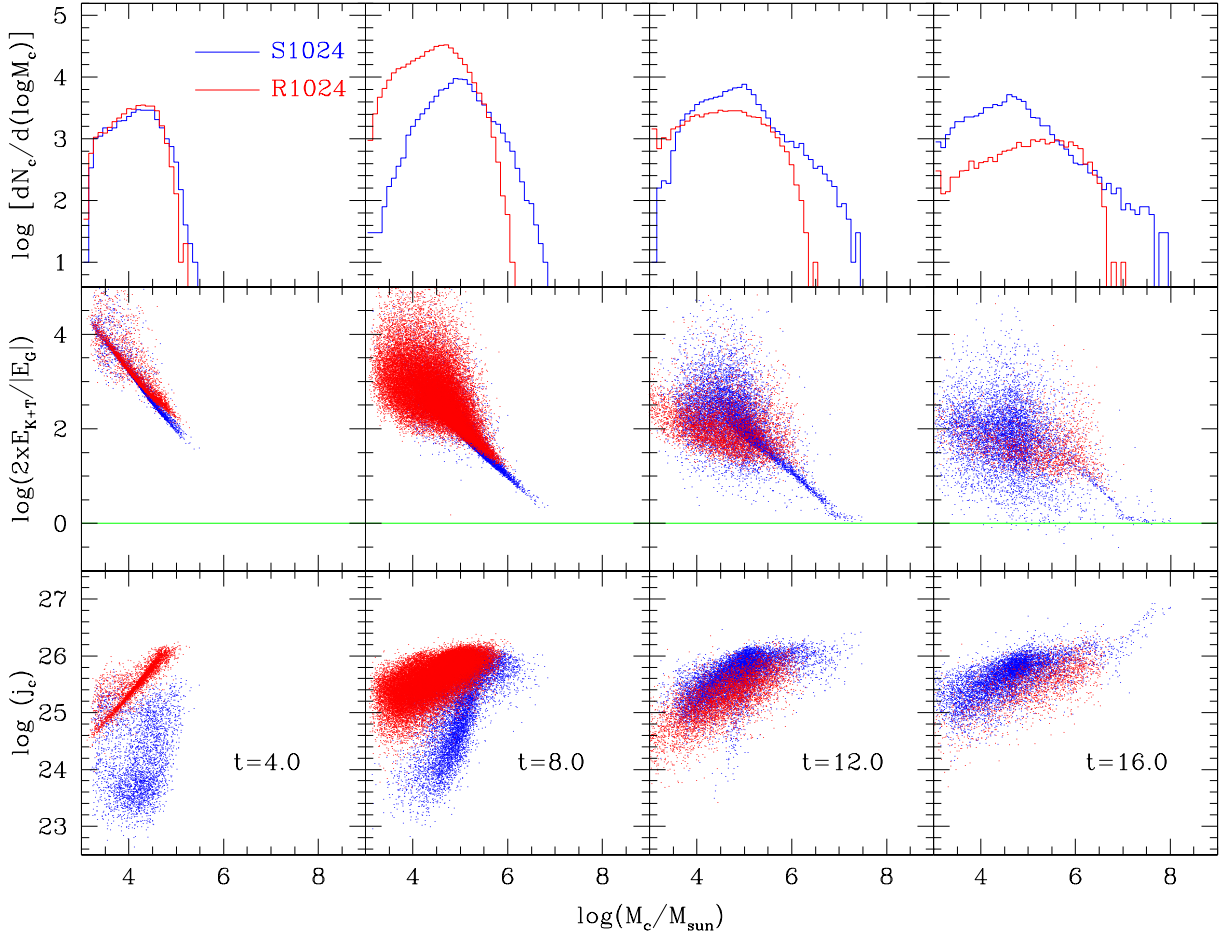}
\figcaption
{Differential number of clouds $dN_c/d(\log M_c)$, energy ratio,
$2 (E_T + E_K) / |E_G|$, and specific angular momentum $j_c$, as a
function of cloud mass $M_c$, at four different times in models
S1024 (blue color) and R1024 (red color).
Here $j_c$ is in cgs units.}
\end{figure}

\end{document}